\documentclass[showpacs,preprintnumbers,prd,nofootinbib,floats,amssymb,floatfix]{revtex4}
\usepackage{amsmath}
\usepackage{graphicx}
\usepackage{amsxtra}
\usepackage{hyperref}
\usepackage{amsmath}
\usepackage{amstext}
\usepackage{blkarray} % Para ponerle indices a las matrices
\begin{document}
\title{The Hamilton-Jacobi analysis  for  higher-order  Chern-Simons  gravity}
\author{Alberto Escalante}  \email{aescalan@ifuap.buap.mx}
\author{ J. Aldair Pantoja-Gonz\'alez}  \email{jpantoja@ifuap.buap.mx}
 \affiliation{  Instituto de F{\'i}sica, Benem\'erita Universidad Aut\'onoma de Puebla. \\
 Apartado Postal J-48 72570, Puebla Pue., M\'exico, }
 \author{D. Vanessa  Castro-Luna} \email{dyana.vcastro@gmail.com}
  \affiliation{Facultad de Ciencias F\'isico Matem\'aticas, Benem\'erita Universidad Aut\'onoma de Puebla.}
\begin{abstract}
Abstract. The Hamilton-Jacobi [$HJ$] analysis for higher-order Chern-Simons gravity is performed. The complete set of $HJ$ Hamiltonians are identified and a fundamental $HJ$ differential is constructed, from which the characteristic equations are obtained. In addition, the symmetries of the theory are identified and the obtained results are compared with other approaches reported in the literature.
\end{abstract}
 \date{\today}
\preprint{}
\maketitle
\section{Introduction}
Nowadays, higher-order theories are an interesting subject in physics. In fact, they are present in modern problems such as dark energy physics \cite{7, 8}, string theory \cite{6}, generalized electrodynamics \cite{2, 3, 4, 5}, noncommutative theory \cite{5a}, and higher-order models of gravity, where those theories are good contender for solving the problem of renormalization in quantum gravity \cite{6a}. There are two kinds of higher-order theories, those with higher-spatial derivatives such as Horava's theory \cite{6a, 7a, 8a}, and those with higher-temporal derivatives such as Weyl theory and theories of gravity with quadratic products of the curvature tensor \cite{9a}. The former is a great proposal as a new theory of gravity at high energy scales. The advantage of this theory is that one can add higher-order spatial derivative terms, the final theory is then expected to be renormalizable. The latter, is also an alternative for quantum gravity due to it have a dimensionless coupling constant which ensures renormalizability, these facts are studied within the so-called $f(R)$  theories \cite{9, 10}. \\
It is important to remark that the theories commented above are singular, and its study is not an easy task to perform. In this respect, we have at hand two formalisms for studying  singular higher-order time derivative theories, the so-called Ostrogradski-Dirac [OD] formulation \cite{11, 12a} and the Hamilton-Jacobi [HJ] method \cite{F17, F18, F19, F20, F21a}. The OD framework is based on  the choice of the fields and their temporal derivatives as canonical variables, thus,  we have an extension of the phase space; the identification of the constraints is then performed as in the Dirac fomalism  is  done \cite{11}. However,  the identification of the constraints into first class or second class is a difficult task because the final  constraints in general,  do not have  a simple structure. On the other hand, the $HJ$ method is based on the identification of the constraints through the null vectors of the  theory, which in this framework are called Hamiltonians. These Hamiltonians can be either involutive or non-involutive; the involutive ones are  used for constructing a fundamental differential. The noninvolutives are removed through the introduction of the  generalized HJ brackets. From the fundamental differential  we can obatain  the characteristic equations, the gauge symmetries,  and the identification of symmetries is  more economical than OD  \cite{22a, 23a, 12, 13}. \\
With all  above, in this paper by using the HJ framework the Chern-Simons gravity theory will be analyzed. In fact, Chern-Simons gravity expressed in terms of a perturbation around the Minkowski background is a higher-order time-derivative theory and it is a good model for applying the $HJ$ scheme \cite{15a}. We are interested in developing the $HJ$ scheme because it is a great alternative for studying gauge theories with higher-order derivatives. In this respect, usually when the OD framework is applied the constraints are fixed by hand, and a clear way for identifying the constraints has not been developed \cite{15a, 16, 17}. On the other hand, in the $HJ$ scheme we can find the correct structure of the Hamiltonians without fixing them by hand, the Hamiltonians are presented with the correct structure and this fact allows us to construct  the fundamental differential. The $HJ$ scheme has been applied generally in ordinary gauge theories and we think that it will be useful for analyzing gauge theories with higher-derivatives. In fact, we will introduce additional fields to reduce the problem to a first-order time derivative one. Consecuently, non-physical degrees of freedom are present and they will be identified as  non-involutive Hamiltonians; at the end, these will be eliminated by introducing  the generalized $HJ$ brackets. In this manner, we will report an alternative development to those reported in the literature where the standard OD framework has been developed \cite{15a}. \\
The paper is organized as follows. In Section II we develop the $HJ$ analysis for the higher-order Chern-Simons gravity. By means of null vectors all Hamiltonians are identified, then we construct a fundamental differential, where the characteristics equations and the symmetries of the theory are found. Finally, we present the  conclusions.
%--------------------------------------------------------------------------
\section{The Hamilton-Jacobi analysis}
We start with the standard linear  form of the Chern-Simons Lagrangian density  given by \cite{22a, 15a}  
\begin{equation}
    L_{CS}=\frac{1}{2}\epsilon^{\lambda \mu \nu} (\partial_\sigma h^{\rho}{}_{\lambda} \partial_\rho \partial_\mu h^{\sigma}{}_{\nu}-\partial_\sigma h^{\rho}{}_{\lambda} \partial^\sigma \partial_\mu h_{\rho \nu}),
\label{eq:lcs}    
\end{equation}
here, spacetime indices are represented by  the greek alphabet $\alpha, \beta = 0, 1, 2$ and space indices  by the latin   $i, j, k =1, 2$, $h_{\mu \nu}$ is the perturbation of the metric around the flat spacetime geometry and the following signature  $\eta_{\mu \nu}= (-1, 1, 1)$ is used. By performing the $2+1$ decomposition, we can write the action as
\begin{equation}
\begin{aligned}
    L_{CS}= {} &  \epsilon^{ij}(\partial^k h_{0j}\Ddot{h}_{ki} + \partial_j h^{k}{}_{0}\Ddot{h}_{ki} - \frac{1}{2} \dot{h}^{k}{}_{j}\Ddot{h}_{ki} + \partial_j \partial^k h_{00} \dot{h}_{ki}+                \partial^k \partial_i h_{0j} \dot{h}_{k0} \\ 
               & + \frac{1}{2} \nabla^2 h_{0j} \dot{h}_{0i} - \frac{1}{2} \nabla^2 h^{k}{}_{j} \dot{h}_{ki} + \frac{1}{2} \partial_k \partial_l h^{l}{}_{j} \dot{h}^{k}{}_{i} + \nabla^2 h_{00} \partial_i h_{0j} \\
               & - \nabla^2 h^{k}{}_{0} \partial_i h_{kj} - \partial^l \partial_i h^{k}_{j} \partial_k h_{l0}).
\end{aligned}
\label{eq:cssep}
\end{equation}
We can observe that the Lagrangian is a higher-order derivative theory, so the standard approach would be to use  the OD framework, however, we have commented above the interest to develop an alternative HJ analysis. It is worth commenting that our results are new and are not reported in the literature.  With this aim, we need to rewrite the Lagrangian by introducing the following variables 
\begin{equation}
    h_{\mu \nu} =  \xi_{\mu \nu}, 
 \quad \quad 
    v_{\mu \nu} = \dot h_{\mu \nu},
\label{eq:var}
\end{equation}
\begin{equation}
\begin{aligned}
    L_{CS}= {} &  \epsilon^{ij}(\partial^k \xi_{0j} \dot{v}_{ki} + \partial_j \xi^{k}{}_{0} \dot{v}_{ki} - \frac{1}{2} v^{k}{}_{j} \dot{v}_{ki} + \partial_j \partial^k \xi_{00} v_{ki} + \partial^k \partial_i \xi_{0j} v_{k0} \\ 
               & + \frac{1}{2} \nabla^2 \xi_{0j} v_{0i} - \frac{1}{2} \nabla^2 \xi^{k}{}_{j} v_{ki} + \frac{1}{2} \partial^k \partial_l \xi^{l}{}_{j} v_{ki} + \nabla^2 \xi_{00} \partial_i \xi_{0j} \\
               & - \nabla^2 \xi^{k}{}_{0} \partial_i \xi_{kj} - \partial^l \partial_i h^{k}{}_{j} \partial_k \xi_{l0}) + \psi^{\alpha \beta} (v_{\alpha \beta}- \dot{\xi}_{\alpha \beta}), 
\end{aligned}
\label{eq:lagcst}    
\end{equation}
where the $\psi's$  are Lagrange multipliers. We can observe that the theory is now linear in the temporal derivatives and we can apply the $HJ$ analysis. From the definition of the momenta 
\begin{eqnarray*}
P^{\mu}=\frac{\partial \mathcal{L}}{\partial \dot{Q}_{\mu}},
\end{eqnarray*}
where $Q_{\mu} = (\xi_{00}, \xi_{0i}, \xi_{ij} v_{00}, v_{0i}, v_{ij},  \psi^{00}, \psi^{0i}, \psi^{ij})$ are the canonical variables and $P^{\mu} = (\pi^{00}, \pi^{0i}, \pi^{ij}, \tilde{\pi}^{00}, \tilde{\pi}^{0i}, \tilde{\pi}^{ij},  p_{00}, p_{0i}, p_{ij})$ their corresponding momenta, we find the following Hamiltonians  \cite{F17,  F18, F19, F20, F21a, 22a, 23a, 12, 13}
\begin{eqnarray}
\Omega_{}^{00} &=& \pi^{00} + \psi^{00} = 0, \nonumber \\
\Omega_{}^{0i} &=& \pi^{0i} + \psi^{0i} = 0, \nonumber \\
\Omega_{}^{ij} &=& \pi^{ij} + \psi^{ij} = 0, \nonumber \\
\tilde{\Omega}_{}^{00} &=& \tilde{\pi}^{00} = 0, \nonumber \\
\tilde{\Omega}_{}^{0i} &=& \tilde{\pi}^{0i} = 0, \nonumber \\
\tilde{\Omega}_{}^{ij} &=& \tilde{\pi}^{ij} - \frac{1}{2}\epsilon^{i l}\partial^j \xi_{0l}- \frac{1}{2}\epsilon^{j l}\partial^i \xi_{0l} - \frac{1}{2}\epsilon^{i l}\partial_l \xi^{{j}}_0 - \frac{1}{2}\epsilon^{j l}\partial_l \xi{^{i}}_0 + \frac{1}{4} \epsilon^{il}v{^{j}}_l+ \frac{1}{4} \epsilon^{jl}v^{{i}}_l= 0, \nonumber \\
\Sigma_{}^{00} &=& p_{00} = 0, \nonumber \\
\Sigma_{}^{0i} &=& p_{0i} = 0,  \nonumber \\
\Sigma_{}^{ij} &=& p_{ij} = 0,
\label{const}
\end{eqnarray}
and the canonical Hamiltonian, given by 
\begin{eqnarray}
\mathcal{H} &=& \dot{\xi}_{\alpha \beta}\pi^{\alpha \beta} + \dot{v}_{\alpha \beta}\tilde{\pi}^{\alpha \beta} + \dot{\psi}_{\alpha \beta}p^{\alpha \beta} - L_{CS} \nonumber  \\[5pt]
 &=&  \epsilon^{ij} \Big( - \partial_j \partial^k \xi_{00}v_{ki} - \partial^k\partial_i \xi_{0j} v_{k0} - \frac{\nabla^2}{2}\xi_{0j} v_{0i} - \frac{\nabla^2}{2} \xi{^k}_{j}v_{ki} - \frac{1}{2} \partial_k \partial_l \xi{^l}_j v{^k}_i - \nabla^2 \xi_{00} \partial_i \xi_{0j}   \nonumber \\
 &+& \nabla^2  \xi{^{k}}_0 \partial_i \xi_{kj}+ \partial^l \partial_i \xi{^k}_j \partial_k \xi_{l0} \Big) - v_{00}\psi^{00} +v_{0i}(\pi^{0i}- 2\psi^{0i}) - v_{ij}\psi^{ij}.       
\end{eqnarray}
With the Hamiltonians identified, we construct the fundamental differential, which describes the evolution
of any function, say $F$, on the phase space \cite{F17, F18, F19, F20, F21a, 12, 13}
\begin{eqnarray}
dF &=& \int \Big[ \{F \;,\; \mathcal{H}\} dt^{0} + \{ F\;,\;\Omega_{}^{00}\} d \omega^{1}_{00} + \{ F\;,\;\Omega_{}^{0i}\} d \omega^{1}_{0i}+ \{F \;,\; \Omega_{}^{ij}\} d \omega^{1}_{ij} + \{F \;,\; \tilde{\Omega}_{}^{00} \} d \omega^{2}_{00} \nonumber \\
&+& \{F \;,\; \tilde{\Omega}_{}^{0i} \} d \omega^{2}_{0i} + \{F \;,\; \tilde{\Omega}_{}^{ij} \} d \omega^{2}_{ij}  
+  \{F \;,\; \Sigma_{}^{00} \} d \omega^{3}_{00} + \{F \;,\; \Sigma_{}^{0i} \} d \omega^{3}_{0i} +\{F \;,\; \Sigma_{}^{ij} \} d \omega^{3}_{ij} \Big] d^{2}y,
\end{eqnarray}
where $ \omega^{1}_{00}, \omega^{1}_{0i}, \omega^{1}_{ij} \omega^{2}_{00},  \omega^{2}_{0 i}, \omega^{2}_{ij},  \omega^{3}_{00},   \omega^{3}_{0i},   \omega^{3}_{ij} $ are parameters associated with the Hamiltonians. Now,  we will  identify  the Hamiltonians into involutive and non-involutive. The  involutive Hamiltonians are those whose Poisson brackets with all Hamiltonians, including themselves, vanish; otherwise, they are called non-involutive. In this manner, the non-zero Poisson algebra between all Hamiltonians (\ref{const}) is given by 
\begin{eqnarray}
\{\Omega_{}^{00}(x) \;,\; \Sigma_{00}(y)\} &=& \delta^2(x-y),  \nonumber \\
\{\Omega_{}^{0i}(x) \;,\; \Sigma_{0j}(y)\} &=& \delta^i{_{j}} \delta^2(x-y),  \nonumber \\
\{\Omega_{}^{ij}(x) \;,\; \Sigma_{kl}(y)\} &=&\frac{1}{2}(\delta^i{_{k}}\delta^j{_{l}}+ \delta^i{_{l} }\delta^j{_{k}}) \delta^{2}(x-y),  \nonumber  \\
\{\Omega_{}^{0i}(x) \;,\; \tilde{\Omega}_{}^{kl}(y)\} &=& -\frac{1}{4}\epsilon^{ki} \partial^l\delta^2(x-y) -\frac{1}{4}\epsilon^{li} \partial^k\delta^2(x-y) -\frac{1}{4}\epsilon^{kj} \eta^{il}\partial_j\delta^2(x-y) -\frac{1}{4}\epsilon^{lj} \eta^{ik}\partial_j\delta^2(x-y),  \nonumber \\
\{\tilde{\Omega}_{}^{ij}(x) \;,\; \tilde{\Omega}_{}^{kl}(y)\} &=& \frac{1}{4} (\epsilon^{jk} \eta^{il}+ \epsilon^{ik} \eta^{jl} + \epsilon^{jl} \eta^{ik} + \epsilon^{il} \eta^{kj}     )\delta^{2}(x-y) 
,\end{eqnarray}
hence, we observe that the  Hamiltonians  $\Omega_{}^{00}, \Omega_{}^{0i}, \Omega_{}^{ij}, \tilde{\Omega}_{}^{ij}, \Sigma_{00}, \Sigma_{0i}, \Sigma_{ij}$ are non-involutives. This is expected because these Hamiltonians are related to the unphysical variables $\psi^{\alpha \beta}$, at the end,   these Hamiltonians will be removed by introducing the generalized HJ bracktes. Furthermore,  Hamiltonians $ \tilde{\Omega}_{}^{00},  \tilde{\Omega}_{}^{0i}$ are involutives. Now, we need remove all non-involutives Hamiltonians, for this step we calculate the matrix composed of the  Poisson brackets between all non-involutives Hamiltonians, namely $\Delta_{a b}$,  this is 
\begin{eqnarray}
\Delta_{a b}=
\begin{pmatrix}
0 & 0 & 0 & 0& 1& 0 & 0& \\[5pt]
0 & 0 &  & \Gamma^{i, kl} & 0 & \eta^{i j} & 0&\\[5pt]
0 & 0& 0 & 0 & 0 & 0 & \frac{1}{2}(\delta^i{_{k}}\delta^j{_{l}}+ \delta^i{_{l} }\delta^j{_{k}})&\\[5pt]
0& - \Gamma^{kl, i}& 0 & \Lambda^{ij, kl}  & 0 & 0& 0&\\[5pt]
-1 & 0 & 0 & 0 & 0 & 0& 0& \\[5pt]
0 & -\eta^{i j} & 0 & 0 & 0 & 0& 0& \\[5pt]
0& 0 & \frac{1}{2}(\delta^i{_{k}}\delta^j{_{l}}+ \delta^i{_{l} }\delta^j{_{k}}) & 0 & 0 & 0& 0& \\[5pt]
\end{pmatrix} \delta^2(x-y) 
\label{matrix}
,\end{eqnarray}
where we have called 
\begin{eqnarray}
\Gamma^{i, kl} &\equiv& -\frac{1}{4}\epsilon^{ki} \partial^l -\frac{1}{4}\epsilon^{li} \partial^k-\frac{1}{4}\epsilon^{kj} \eta^{il}\partial_j-\frac{1}{4}\epsilon^{lj} \eta^{ik}\partial_j,  \nonumber \\
\Lambda^{ij, kl} &\equiv&   \frac{1}{4} (\epsilon^{jk} \eta^{il}+ \epsilon^{ik} \eta^{jl} + \epsilon^{jl} \eta^{ik} + \epsilon^{il} \eta^{kj}).
\end{eqnarray} 
We can observe that (\ref{matrix}) is not invertible, which means that the Hamiltonians are not independent. In fact, there are null vectors given by $\vec{v}= (0, 0, 0, \eta_{kl} \varpi,0, 0, 0 )$, where $\varpi$  is an arbitrary function. From contraction of the null vectors with the Hamiltonians (\ref{const})  we identify the following new Hamiltonian 
\begin{equation}
\tilde{\Omega} :\tilde{\pi}^{i}{_{i}}=0. 
\end{equation}
This Hamiltonian is involutive, then,  this implies that the new non-involutives Hamiltonians are now 
\begin{eqnarray}
\Omega_{}^{00} &=& \pi^{00} + \psi^{00} = 0, \nonumber \\
\Omega_{}^{0i} &=& \pi^{0i} + \psi^{0i} = 0, \nonumber \\
\Omega_{}^{ij} &=& \pi^{ij} + \psi^{ij} = 0, \nonumber \\
\tilde{\Omega}_{}^{11} &=& \tilde{\pi}^{11} -\partial^1\xi_{02}- \partial^2\xi_{10}+ \frac{1}{2}v^1{_{2}}= 0, \nonumber \\
\tilde{\Omega}_{}^{12} &=& \tilde{\pi}^{12} -\partial^2\xi_{02}- \partial^1\xi_{10}+ \frac{1}{4}v{_{22}}+ \frac{1}{4}v{_{11}}= 0, \nonumber \\
\Sigma_{}^{00} &=& p_{00} = 0, \nonumber \\
\Sigma_{}^{0i} &=& p_{0i} = 0,  \nonumber \\
\Sigma_{}^{ij} &=& p_{ij} = 0, 
\label{const2}
\end{eqnarray}
with the new non-involutives Hamiltonians, we calculate the matrix whose entries are   the  Poisson brackets between them, it is given by 
\begin{eqnarray}
{\Delta{_2}}{_{a b}}=
\begin{pmatrix}
0 & 0 & 0 & 0& 0& 0 & 1& 0& 0& 0  \\
0 & 0 & 0 & 0& -\frac{1}{2}\partial_2& \frac{1}{2}\partial_1& 0& 1& 0& 0\\
0 & 0& 0 & 0 & -\frac{1}{2}\partial_1 & -\frac{1}{2}\partial_2&& 0& 1& 0\\
0& & 0 &  & 0 & 0& 0& 0& 0& \frac{1}{2}(\delta^i{_{k}}\delta^j{_{l}}+ \delta^i{_{l} }\delta^j{_{k}}) \\
 0& \frac{1}{2}\partial_2 & \frac{1}{2}\partial_1 & 0 & 0 & \frac{1}{2}& 0& 0& 0& 0 \\
0 &  -\frac{1}{2}\partial_1& \frac{1}{2}\partial_2& 0 & - \frac{1}{2} & 0& 0& 0& 0& 0\\
-1& 0 & 0 & 0 & 0 & 0& 0& 0& 0& 0 \\
0 &  -1& 0 & 0 & 0 & 0& 0& 0& 0& 0\\
0 &  0& -1 & 0 & 0 & 0& 0& 0& 0& 0\\
0 & 0 & 0 &-\frac{1}{2}(\delta^i{_{k}}\delta^j{_{l}}+ \delta^i{_{l} }\delta^j{_{k}})  & 0 & 0& 0& 0& 0& 0\\
\end{pmatrix} \delta^2(x-y), \nonumber  \\
\label{matrix2}
\end{eqnarray}
the inverse of (\ref{matrix2}) is given by 
\begin{eqnarray}
({\Delta{_2}{^{a b}}})^{-1}=
\begin{pmatrix}
0 & 0 & 0 & 0& 0& 0 & 1& 0& 0& 0  \\
0 & 0 & 0 & 0& -\frac{1}{2}\partial_2& \frac{1}{2}\partial_1& 0& 1& 0& 0\\
0 & 0& 0 & 0 & -\frac{1}{2}\partial_1 & -\frac{1}{2}\partial_2&0& 0& 1& 0\\
0&0 & 0 &0  & 0 & 0& 0& 0& 0& \frac{1}{2}(\delta^i{_{k}}\delta^j{_{l}}+ \delta^i{_{l} }\delta^j{_{k}}) \\
 0& \frac{1}{2}\partial_2 & \frac{1}{2}\partial_1 & 0 & 0 & \frac{1}{2}& 0& 0& 0& 0 \\
0 &  -\frac{1}{2}\partial_1& \frac{1}{2}\partial_2& 0 & - \frac{1}{2} & 0& 0& 0& 0& 0\\
-1& 0 & 0 & 0 & 0 & 0& 0& 0& 0& 0 \\
0 &  -1& 0 & 0 & 0 & 0& 0& 0& 0& 0\\
0 &  0& -1 & 0 & 0 & 0& 0& 0& 0& 0\\
0 & 0 & 0 &-\frac{1}{2}(\delta^i{_{k}}\delta^j{_{l}}+ \delta^i{_{l} }\delta^j{_{k}})  & 0 & 0& 0& 0& 0& 0\\
\end{pmatrix} \delta^2(x-y), \nonumber  \\
\label{matrix3}
\end{eqnarray}
With this inverse matrix  we introduce the $HJ$ generalized brackets, defined as
\begin{eqnarray}
\{A(x), B(x^{\prime})\}^* = \{A(x), B(x^{\prime})\} - \int\int  \{A(x), \zeta^{a}(y)\}( {\Delta_2}{^{a b}})^{{-1}}(y, z) \{\zeta^{b}(z), B(x^{\prime})\} \; d^{2}y \; d^{2}z
\label{gbra} 
,\end{eqnarray}
where $\zeta^{a}$  represent the non-involutive Hamiltonians. Hence, with the generalized brackets (\ref{gbra}) we  calculate those  between the phase space variables, these are 
\begin{eqnarray}
&\left\lbrace \xi_{00},\pi^{00} \right\rbrace^{*}& = \delta^{2}(x-y), \nonumber \\
&\left\lbrace \xi_{0i},\pi^{0l} \right\rbrace^{*}& = \frac{1}{2}\delta_{l}{}^{i}\delta^{2}(x-y), \nonumber \\
&\left\lbrace \xi_{ij},\pi^{lm} \right\rbrace^{*}& = \frac{1}{2} \left( \delta_{i}{}^{l}\delta_{j}{}^{m} + \delta_{i}{}^{m}\delta_{j}{}^{l}\right) \delta^{2}(x-y), \nonumber \\
&\left\lbrace \pi^{0i},\pi^{0l} \right\rbrace^{*}& = \frac{1}{2} \epsilon^{il}\nabla^{2} \delta^{2}(x-y), \nonumber \\
&\left\lbrace \pi^{0i},v_{lm} \right\rbrace^{*}& = -\frac{1}{2} \left( \delta_{l}{}^{1}\delta_{m}{}^{2} + \delta_{l}{}^{2}\delta_{m}{}^{1} \right) \left( \epsilon^{1i}\partial^{1} +  \eta^{1i}\partial^{2} \right) \delta^{2}(x-y) \nonumber \\
&&+ \frac{1}{2}\delta_{l}{}^{1}\delta_{m}{}^{1} \left( \epsilon^{1i}\partial^{2} +  \epsilon^{2i}\partial^{1} + \eta^{2i}\partial^{2} - \eta^{1i}\partial^{1} \right) \delta^{2}(x-y), 
\end{eqnarray}
We can observe that these generalized brackets coincide with those reported in \cite{22a} where alternative methods were used.  With the introduction of the generalized brackets, the non-involutives Hamiltonians can be removed, then, the fundamental differential will be 
\begin{eqnarray}
dF &=& \int \Big[ \{F \;,\; \mathcal{H}\}^* dt^{0} + \{F \;,\; \tilde{\Omega}_{}^{00} \}^* d \omega^{2}_{00} 
+  \{F \;,\; \tilde{\Omega}_{}^{0i} \}^* d \omega^{2}_{0i} +\{F \;,\; \tilde{\Omega}_{}^{} \}^* d \omega^{2}_{}   \Big] d^{2}y,
\end{eqnarray}
where $\tilde{\Omega}^{00}, \tilde{\Omega}^{0i}, \tilde{\Omega} $ are involutive ones. Once the generalized brackets are introduced, we could make the substitution of the fields $\psi$ by the momenta $\pi$,   and the canonical Hamiltonian takes the form
\begin{eqnarray}
\mathcal{H} &=& \epsilon^{ij} \Big( - \partial_j \partial^k \xi_{00}v_{ki} - \partial^k\partial_i \xi_{0j} v_{k0} - \frac{\nabla^2 \xi_{0j}}{2} v_{0i} - \frac{\nabla^2 \xi{^k}_{j}}{2} v_{ki} - \frac{1}{2} \partial_k \partial_l \xi{^l}_j v{^k}_i - \nabla^2 \xi_{00} \partial_i \xi_{0j}   \nonumber \\
 &+& \nabla^2  \xi{^{k}}_0 \partial_i \xi_{kj}+ \partial^l \partial_i \xi{^k}_j \partial_k \xi_{l0} \Big) - v_{00}\pi^{00} +2v_{0i}\pi^{0i} - v_{ij}\pi^{ij}.       
\end{eqnarray}
From Frobenius integrability conditions, which ensures  the  integrability of the system, the following Hamiltonians emerge
\begin{eqnarray}
d \tilde{\Omega}^{00}&=&\int \Big[ \{ \tilde{\Omega}^{00} \;,\; \mathcal{H} \}^* dt^{0} + \{ \tilde{\Omega}^{00} \;,\; \tilde{\Omega}^{00} \}^* d \omega^{2}_{00} 
+  \{ \tilde{\Omega}^{00} \;,\; \tilde{\Omega}^{0i} \}^* d \omega^{2}_{0i} +\{ \tilde{\Omega}^{00} \;,\; \tilde{\Omega} \} ^*d \omega^{2}  \Big] d^{2}y =0 \nonumber  \\ 
&\rightarrow& \tilde{\Omega}_{2}^{00}  \equiv   \pi^{00}=0, \\
d \tilde{\Omega}^{0i}&=&\int \Big[ \{ \tilde{\Omega}^{0i} \;,\; \mathcal{H} \}^* dt^{0} + \{ \tilde{\Omega}^{0i} \;,\; \tilde{\Omega}^{00} \}^* d \omega^{2}_{00} 
+  \{ \tilde{\Omega}^{0i} \;,\; \tilde{\Omega}^{0i} \}^* d \omega^{2}_{0i} +\{ \tilde{\Omega}^{0i} \;,\; \tilde{\Omega} \}^* d \omega^{2}  \Big] d^{2}y =0 \nonumber  \\ 
&\rightarrow& \tilde{\Omega}_{2}^{0i}  \equiv   \pi^{0i}- \frac{1}{2}\epsilon^{jl} \partial^i \partial_j \xi_{0l} - \frac{1}{4} \epsilon^{ij} \nabla^2 \xi_{0j}=0, \\
d \tilde{\Omega}&=&\int \Big[ \{ \tilde{\Omega} \;,\; \mathcal{H} \}^* dt^{0} + \{ \tilde{\Omega} \;,\; \tilde{\Omega}^{00} \}^* d \omega^{2}_{00} 
+  \{ \tilde{\Omega} \;,\; \tilde{\Omega}^{0i} \}^* d \omega^{2}_{0i} +\{ \tilde{\Omega} \;,\; \tilde{\Omega} \}^* d \omega^{2}  \Big] d^{2}y =0 \nonumber  \\ 
&\rightarrow& \tilde{\Omega}_{2}  \equiv   \pi^{i}{_{i}}-\frac{1}{2} \epsilon^{ij} \partial_i \partial^l \xi_{lj}=0, 
\label{invo3}
\end{eqnarray}
 after a long algebraic work, we can  observe that  $ \tilde{\Omega}_{2}^{00},  \tilde{\Omega}_{2}^{0i},  \tilde{\Omega}_{2}  $ are   involutive Hamiltonians because the Poisson brackets with   $\tilde{\Omega}^{00}, \tilde{\Omega}^{0i}, \tilde{\Omega} $ and themselfves vanishes. Since the new Hamiltonians are involutives, we will  add them to the fundamental differential,  then its integrability will be calculated. From integrability conditions we obtain
 \begin{eqnarray}
d\tilde{\Omega}_{2}^{00} &=& \int \Big[ \{\tilde{\Omega}_{2}^{00} \;,\; \mathcal{H}\}^* dt^{0} + \{\tilde{\Omega}_{2}^{00} \;,\; \tilde{\Omega}_{}^{00} \}^* d {\omega_{2}}_{00} 
+  \{\tilde{\Omega}_{2}^{00} \;,\; \tilde{\Omega}_{}^{0i} \}^* {d \omega_{2}}_{0i} +\{\tilde{\Omega}_{2}^{00} \;,\; \tilde{\Omega}_{}^{} \}^* d{ \omega_{2}}_{}   \nonumber \\
&+&  \{\tilde{\Omega}_{2}^{00} \;,\; \tilde{\Omega}_{2}^{00} \}^* d \tilde{\omega}_{00} 
+  \{\tilde{\Omega}_{2}^{00} \;,\; \tilde{\Omega}_{2}^{0i} \}^* d \tilde{\omega}_{ 0i} +\{\tilde{\Omega}_{2}^{00} \;,\; \tilde{\Omega}_{2}^{} \}^* d \tilde{\omega}_{} \Big] d^{2}y, \nonumber \\
&\rightarrow& \tilde{\Omega}^{00}_{3}  \equiv   \epsilon^{ij}\partial_j \partial^k v_{ki} + \epsilon^{ij} \nabla^2 \partial_i \xi_{0j}=0, \nonumber \\
d\tilde{\Omega}_{2}^{0i} &=& \int \Big[ \{\tilde{\Omega}_{2}^{0i} \;,\; \mathcal{H}\}^* dt^{0} + \{\tilde{\Omega}_{2}^{0i} \;,\; \tilde{\Omega}_{}^{00} \}^* d {\omega_{2}}_{00} 
+  \{\tilde{\Omega}_{2}^{0i} \;,\; \tilde{\Omega}_{}^{0i} \}^* {d \omega_{2}}_{0i} +\{\tilde{\Omega}_{2}^{0i} \;,\; \tilde{\Omega}_{}^{} \}^* d{ \omega_{2}}_{}   \nonumber \\
&+&  \{\tilde{\Omega}_{2}^{0i} \;,\; \tilde{\Omega}_{2}^{00} \}^* d \tilde{\omega}_{00} 
+  \{\tilde{\Omega}_{2}^{0i} \;,\; \tilde{\Omega}_{2}^{0i} \}^* d \tilde{\omega}_{ 0i} +\{\tilde{\Omega}_{2}^{0i} \;,\; \tilde{\Omega}_{2}^{} \}^* d \tilde{\omega}_{} \Big] d^{2}y, \nonumber \\
&\rightarrow& \tilde{\Omega}^{0i}_{3}  \equiv   \partial_j\pi^{ij}- \frac{1}{4} \epsilon^{jl}\nabla^2 \partial_j \xi^{i}{_{l}}- \frac{1}{4} \epsilon^{jl}\partial^i \partial_j \partial_k \xi^{k}{_{l}}=0, \nonumber \\
d\tilde{\Omega}_{2} &=& \int \Big[ \{\tilde{\Omega}_{2} \;,\; \mathcal{H}\}^* dt^{0} + \{\tilde{\Omega}_{2} \;,\; \tilde{\Omega}_{}^{00} \}^* d {\omega_{2}}_{00} 
+  \{\tilde{\Omega}_{2} \;,\; \tilde{\Omega}_{}^{0i} \}^* {d \omega_{2}}_{0i} +\{\tilde{\Omega}_{2} \;,\; \tilde{\Omega}_{}^{} \}^* d{ \omega_{2}}_{}   \nonumber \\
&+&  \{\tilde{\Omega}_{2} \;,\; \tilde{\Omega}_{2}^{00} \}^* d \tilde{\omega}_{00} 
+  \{\tilde{\Omega}_{2} \;,\; \tilde{\Omega}_{2}^{0i} \}^* d \tilde{\omega}_{ 0i} +\{\tilde{\Omega}_{2} \;,\; \tilde{\Omega}_{2}^{} \}^* d \tilde{\omega}_{} \Big] d^{2}y, \nonumber \\
&\rightarrow& \tilde{\Omega}^{00}_{3}=0, 
\end{eqnarray}
the new third generation of Hamiltonians $  \tilde{\Omega}^{00}_{3}, \tilde{\Omega}^{0i}_{3} $  are involutives and from their  integrability we find no further  Hamiltonians. Then the complete fundamental differential  is given by 
\begin{eqnarray}
dF &=& \int \Big[ \{F \;,\; \mathcal{H}\}^* dt^{0} + \{F \;,\; \tilde{\Omega}_{}^{00} \}^* d {\omega_{2}}_{00} 
+  \{F \;,\; \tilde{\Omega}_{}^{0i} \}^* {d \omega_{2}}_{0i} +\{F \;,\; \tilde{\Omega}_{}^{} \}^* d{ \omega_{2}}_{}   \nonumber \\
&+&  \{F \;,\; \tilde{\Omega}_{2}^{00} \}^* d \tilde{\omega}_{00} 
+  \{F \;,\; \tilde{\Omega}_{2}^{0i} \}^* d \tilde{\omega}_{ 0i} +\{F \;,\; \tilde{\Omega}_{2}^{} \}^* d \tilde{\omega}_{} + \{F \;,\; \tilde{\Omega}_{3}^{00} \}^* d {\tilde{\omega}_{300}} \nonumber \\
&+&\{F \;,\; \tilde{\Omega}_{3}^{0i} \}^* d {\tilde{\omega}_{3 0i}} \Big] d^{2}y,
\label{23}
\end{eqnarray}
and the complete set of involutives Hamiltonians are 
\begin{eqnarray}
\tilde{\Omega}_{}^{00} &\equiv& \tilde{\pi}^{00} = 0, \nonumber \\
\tilde{\Omega}_{}^{0i} &\equiv& \tilde{\pi}^{0i} = 0, \nonumber \\
\tilde{\Omega} &\equiv&\tilde{\pi}^{i}{_{i}}=0, \nonumber \\
 \tilde{\Omega}_{2}^{00}  &\equiv &   \pi^{00}=0, \nonumber  \\
 \tilde{\Omega}_{2}^{0i} &\equiv&  \pi^{0i}- \frac{1}{2}\epsilon^{jl} \partial^i \partial_j \xi_{0l} - \frac{1}{4} \epsilon^{ij} \nabla^2 \xi_{0j}=0, \nonumber \\
 \tilde{\Omega}_{2}  &\equiv&   \pi^{i}{_{i}}-\frac{1}{2} \epsilon^{ij} \partial_i \partial^l \xi_{lj}=0, \nonumber \\
 \tilde{\Omega}^{00}_{3}  &\equiv&   \epsilon^{ij}\partial_j \partial^k v_{ki} + \epsilon^{ij} \nabla^2 \partial_i \xi_{0j}=0, \nonumber \\
  \tilde{\Omega}^{0i}_{3}  &\equiv&   \partial_j\pi^{ij}- \frac{1}{4} \epsilon^{jl}\nabla^2 \partial_j \xi^{i}{_{l}}- \frac{1}{4} \epsilon^{jl}\partial^i \partial_j \partial_k \xi^{k}{_{l}}=0,
\end{eqnarray}
we can observe that these Hamiltonians have a smaller structure than those reported in \cite{22a} where alternatives methods were used. In fact, in our final set of Hamiltonians, the noninvolutive ones do not appear any more and therefore the structure is smaller. \\
Futheremore, from the fundamental differential (\ref{23}) we will calculate the characteristic equations for the fields $\xi's$, which will reveal the symmetries of the theory. We find them to be
\begin{eqnarray}
\nonumber d\xi_{00} &=& -v_{00}dt - d\tilde{\omega}_{00}, \\ \nonumber 
d\xi_{0i} &=& v_{0i}dt +\frac{1}{2}d\tilde{\omega}_{0i}, \\ 
d\xi_{ij} &=& v_{ij}dt + \eta_{ij}d\tilde{\omega} -\frac{1}{2} \partial_id\tilde{\omega}_{30j} - \frac{1}{2} \partial_j d\tilde{\omega}_{30i}, 
\label{25}
\end{eqnarray}
It is well known that  the corresponding Hamiltonians $\Omega's$ are  generators  of canonical transformations \cite{F19, F20}. In this manner,  to relate these canonical transformations to the gauge ones we set $dt=0$, obtaining 
\begin{eqnarray}
\nonumber \delta\xi_{00} &=& - \delta \tilde{\omega}_{00}, \\ \nonumber 
\delta\xi_{0i} &=& \frac{1}{2}\delta\tilde{\omega}_{0i}, \\ 
\delta\xi_{ij} &=&  \eta_{ij}\delta\tilde{\omega} -\frac{1}{2} \partial_i\delta\tilde{\omega}_{30j} - \frac{1}{2} \partial_j \delta\tilde{\omega}_{30i}, 
\label{26}
\end{eqnarray}
In order  to identify  the gauge transformations it is necessary to find  the  conditions in which (\ref{26}) acts into the Lagrangian. These conditions are given when  the Lagrangian (\ref{eq:lcs}) is   invariant under these transformations if $\delta L =0$. This will result in relations between the parameters $\tilde{\omega}'s$. The variation of the Lagrangian is given by 
\begin{eqnarray}
\delta L = \int dt\;d^{2}x \epsilon^{\alpha \mu \nu} \left[ \partial^\rho \partial_\rho \partial_\mu \xi^{\beta}{_{\nu}}- \partial^\sigma \partial^\beta \partial_{\mu} \xi_{\sigma \nu}   \right] \delta \xi_{\alpha \beta} = 0, 
\label{varia}
\end{eqnarray}
thus,  the variation  takes the form 
\begin{eqnarray}
\delta L &=& \int dt\;d^{2}x\; \Big\{ \left( \epsilon^{0i\nu} \partial^\rho \partial_\rho\partial_i \xi^{0}{_{\nu}} -     \epsilon^{0i\nu} \partial^\sigma \partial^0\partial_i \xi_{\sigma \nu} \right)\delta \xi_{00} + \left( \epsilon^{0j\nu} \partial^\rho \partial_\rho\partial_j \xi^{i}{_{\nu}} -     \epsilon^{0j\nu} \partial^\sigma \partial^i\partial_j \xi_{\sigma \nu} \right)\delta \xi_{0i} \nonumber \\
&+& \left( \epsilon^{i0\nu} \partial^\rho \partial_\rho\partial_0 \xi^{0}{_{\nu}} -     \epsilon^{i0\nu} \partial^\sigma \partial^0\partial_0 \xi_{\sigma \nu} \right)\delta \xi_{0i} +  \left( \epsilon^{ij\nu} \partial^\rho \partial_\rho\partial_j \xi^{0}{_{\nu}} -     \epsilon^{ij\nu} \partial^\sigma \partial^0\partial_j \xi_{\sigma \nu} \right)\delta \xi_{0i} \nonumber \\
&+& \left( \epsilon^{i0\nu} \partial^\rho \partial_\rho \partial_0 \xi^{j}{_{\nu}} -     \epsilon^{i0\nu} \partial^\sigma \partial^j \partial_0 \xi_{\sigma \nu} \right) \delta \xi_{ij} +\left( \epsilon^{ik\nu} \partial^\rho \partial_\rho\partial_k \xi^{j}{_{\nu}} -     \epsilon^{ik\nu} \partial^\sigma \partial^j\partial_k \xi_{\sigma \nu} \right)\delta \xi_{ij}  \Big\} =0.
\end{eqnarray}
The theory will  be invariant under (\ref{26})  if the  $\omega's$  parameters  obey
\begin{eqnarray}
 \delta \tilde{\omega}_{00} &=& -2\partial_0 \zeta_0,  \nonumber \\ 
   \delta \tilde{\omega}_{0i} &= & 2(\partial_0 \zeta_i+ \partial_i \zeta_0), \nonumber \\ 
 \eta_{ij} \delta \tilde{\omega} &=&  2(\partial_i \zeta_j + \partial_j \zeta_i), \nonumber \\ 
 \delta \tilde{\omega}_{30i} &=& 2\zeta_i 
 \label{29}
\end{eqnarray}
hence, from (\ref{29}) the gauge transformations are given by
\begin{eqnarray}
\delta \xi_{\mu \nu}= \partial_\mu \zeta_\nu + \partial_\nu \zeta_\mu,
\label{gauge}
\end{eqnarray}
these gauge transformations are those reported in \cite{22a}. In this manner, our results extend those reported in the literature. 
%%%%%%%
%%%%%%%5
% \subsection{$HJ$} ends here
\section{Conclusions}
In this paper the HJ framework for CS higher-order theory has been applied. By means of null vectors the complete set of Hamiltonians was obtained. This fact allowed us to identify that the complete structure of the Hamiltonias  is smaller than that reported reported in \cite{22a, 15a} were alternative methods were applied. The generalized HJ brackets were constructed and the generalized differential  was obtained. From the generalized differential the characteristic equations were identified and the gauge transformations  reproduced. We observed that the HJ method is a good alternative for analyzing higher-order theories and it is  more economical than the standard approaches. In this manner, we expect  that the HJ framework can be applied to theories with physical degrees of freedom. However, all these aims are under study and will be reported in forthcoming works \cite{17a}.  

\end{document}